\documentclass[12pt]{article}
\usepackage{amsmath,amssymb,epsfig}

\usepackage{color}
\input{colordvi.tex}
\def\unit{{\relax{\rm 1\kern-.26em I}}}

\makeatletter

\renewcommand\section{\@startsection {section}{1}{\z@}%
                                   {-3.5ex \@plus -1ex \@minus -.2ex}%
                                   {2.3ex \@plus.2ex}%
                                   {\normalfont\large\bfseries}}

\renewcommand\subsection{\@startsection{subsection}{2}{\z@}%
                                     {-3.25ex\@plus -1ex \@minus -.2ex}%
                                     {1.5ex \@plus .2ex}%
                                     {\normalfont\normalsize\bfseries}}

\makeatother

\begin{document}

\baselineskip=18pt  
\numberwithin{equation}{section}  
\allowdisplaybreaks  



%
%


\thispagestyle{empty}

\vspace*{-2cm}
\begin{flushright}
\end{flushright}

\begin{flushright}
\end{flushright}

\begin{center}

\vspace{1.4cm}

\vspace{1cm}
{\bf\Large Baryon as Impurity for Phase Transition} 

\vspace{0.2cm}
{\bf\Large in String Landscape}
\vspace*{1.3cm}

{\bf
Aya Kasai$^{1}$, Yuichiro Nakai$^{2}$ and Yutaka Ookouchi$^{3,1}$} \\
\vspace*{0.5cm}

${ }^{1}${\it Department of Physics, Kyushu University, Fukuoka 810-8581, Japan  }\\
${ }^{2}${\it Department of Physics, Harvard University, Cambridge, MA 02138, USA}\\
${ }^{3}${\it Faculty of Arts and Science, Kyushu University, Fukuoka 819-0395, Japan  }\\

\vspace*{0.5cm}

\end{center}

\vspace{1cm} \centerline{\bf Abstract} \vspace*{0.5cm}

We consider a decay of a false vacuum in flux compactifications of type IIB string theory and
study a catalytic effect for a phase transition induced by a new type of impurities.
We concentrate on the large $N$ dual of a D5-brane/anti-D5-brane system which has a rich vacuum structure. We show that D3-branes wrapping the 3-cycles can form a dibaryon and make a bound state with a monopole. We find that these baryon-like objects can make the lifetime of the metastable vacuum shorter.

\newpage
\setcounter{page}{1} 



\section{Introduction}

Catalytic effects for phase transitions are ubiquitous in nature.
The evolution of the universe (or multiverse) may be affected by their existence in that
the catalysts accelerate transitions between vacua.
In fact, it has been discussed in field theories that a semi-classical vacuum decay occurs from monopoles
or cosmic strings
\cite{Steinhardt:1981ec,Hosotani:1982ii,Yajnik:1986tg}
and applied to phenomenological studies of metastable vacua
\cite{Kumar:2010mv,Hiramatsu:2013yxa,Lee:2013ega}.
Moreover, recent developments of string theory compactifications suggest the existence of the landscape of vacua
\cite{Bousso:2000xa,Kachru:2003aw}. Inspired by this idea, non-supersymmetric vacua have been constructed in type IIB string theory, for example see 
\cite{Kachru:2003aw,Kachru:1999vj,Vafa:2000wi,Kachru:2002gs,Argurio:2006ny,Aganagic:2006ex,YO} (for a review, see \cite{KOOreview}). 
Also in type IIA string theory, non-supersymmetric vacua, which are related to metastable vacua in supersymmetric QCD \cite{Intriligator:2006dd}, were realized in the context of brane configurations \cite{Ooguri:2006bg,Franco:2006ht,Bena:2006rg,Tatar:2006dm}. If the landscape really exists, it is necessary to elucidate how effectively  phase transitions proceed
in order to uncover the cosmic history.
As emphasized in \cite{Kasai:2015exa,Kasai:2015dia}, catalytic effects in string theory may play an important role in this investigation.
It was discussed in Ref.~\cite{Kasai:2015exa} that stringy monopoles can accelerate a decay of a non-supersymmetric vacuum,
whose geometry  is constructed from D5-branes and anti-D5-branes
wrapping homologous and minimal sized 2-spheres inside a Calabi-Yau manifold
\cite{Aganagic:2006ex}.
The magnetic monopoles are realized by D3-branes wrapping a 3-chain between D5/anti-D5-branes.
In the decay process, the D3-branes form a bound state with a domain wall D5-brane wrapping the 3-chain.
It was shown that the stability of the false vacuum depends on the strength of the magnetic flux induced by the dissolving D3-branes. Thus, it would be important to study the catalytic effect in flux compactifications to extract a lesson for the string landscape.

The large $N$ dual picture of this D5-brane/anti-D5-brane system is given by
geometric transition which replaces the wrapped 2-spheres with finite sized 3-spheres where Ramond-Ramond (RR) fluxes are turned on
\cite{Vafa:2000wi,Aganagic:2006ex,Cachazo:2001jy}.
Ref.~\cite{Heckman:2007wk} further analyzed the phase structure of this system.
In addition to classification of the vacua in terms of the number of the flux lines through the 3-spheres,
there are many metastable states with the same number of the flux lines, which can be interpreted as $N$ confining vacua present in the $\mathcal{N} = 1$ $SU(N)$ pure Yang-Mills theory in the limit where branes and anit-branes are infinitely separated. These $N$ vacua are degenerate at the leading order in SUSY breaking but higher order corrections,
calculated by an auxiliary matrix model
\cite{Dijkgraaf}, lift this degeneracy.
Then, transitions between the vacua proceed by wrapping
D5-branes over the 3-cycles in the manifold like brane/flux annihilation
\cite{Kachru:2002gs} (See also Ref.~\cite{Frey:2003dm,Brown:2009yb,Bena:2014jaa,Danielsson:2014yga,Gautason:2015tla}).\footnote{
Ref.~\cite{Kachru:2002gs} discussed NS5-brane mediated annihilation between anti-D3 branes and an NSNS 3-form flux
in the Klebanov-Strassler geometry
\cite{Klebanov:2000hb}.}
As in the case of the open string picture, a D5-brane wrapping a compact 3-cycle connecting the 3-spheres
is a domain wall solution for a vacuum decay reducing the number of the RR flux lines.
Furthermore, D5-branes wrapping the 3-spheres at tips of the conifold points are domain walls separating the multiple confining vacua.

In this paper, we consider the large $N$ dual picture of the D5-brane/anti-D5-brane system and
study a catalytic effect for a transition between the vacua from a new type of impurities, a bound state of a dibaryon
\cite{dibaryon} with a monopole.
The existence of a baryon vertex was firstly proposed by Ref.~\cite{Witten:1998xy} in the context of the AdS/CFT correspondence
\cite{Aharony:1999ti}.
While a stringy monopole is realized by wrapping a D3-brane over a compact 3-cycle as discussed in Ref.~\cite{Kasai:2015exa},
the presence of background RR fluxes in a 3-sphere induces
the charge of a gauge field on a wrapped D3-brane world-volume.
Since the total charge of a gauge field has to vanish in a closed cycle,
the D3-brane is attached by fundamental strings, which provide other sources of the gauge charge,
and a dibaryon is formed to neutralize the charge. Due to the presence of fundamental strings, connecting the D3-branes wrapping the different 3-cycles,
dissolving the fundamental string into the D3-brane corresponding to a monopole makes the total energy lower. During a phase transition, the baryon forms a bound state with a domain wall
and induces the magnetic flux on the wall as in the case of Ref.~\cite{Kasai:2015exa}. By this impurity, the catalytic effect is maximized when the domain walls for the two types of phase transitions, reducing the RR flux lines and moving into an adjacent confining vacuum, expand from a same point in our spacetime.
It is shown that the dibaryon can accelerate the transitions
and sometimes threaten the vacuum stability itself.

The rest of the paper is organized as follows.
In section $2$, we will present our geometrical setup.
The discussion of Ref.~\cite{Aganagic:2006ex,Heckman:2007wk} on the large $N$ dual of a brane/anti-brane system
is reviewed.
We discuss scenarios of decays of false vacua in the rich phase structure.
In section $3$, 
additional D3-branes wrapping the 3-cycles will be introduced.
Due to the nonzero RR fluxes, they are connected by fundamental strings
and are interpreted as baryon-like objects, dibaryons.
In section $4$, we will analyze a catalytic effect on a vacuum decay from the dibaryon.
We first look at different phase transitions one by one
and focus on the case where the multiple transitions occur at a same point in our spacetime
and the presence of the dibaryon is the most effective.
We analyze the stability of the false vacuum
and estimate the decay rate in the metastable case.
In section $5$, we will conclude discussions and comment on possible future directions.

\section{The brane/anti-brane system at large $N$}

In this section, we introduce our geometrical setup without impurities presented by
Ref.~\cite{Aganagic:2006ex,Heckman:2007wk}
and review the discussion on the large $N$ duality between the open and closed string pictures via geometric transition
and the phase structure of the system.
The closed string picture is the starting point of later discussions with baryon impurities.
Metastability and decays of false vacua in this geometry are also reviewed.

\subsection{Review of geometrically induced metastable vacua }

Consider a noncompact Calabi-Yau threefold with the following defining equation \cite{Cachazo:2001jy}:
\begin{equation}
\begin{split}
\\[-2.5ex]
w z  = y^2 + W'(x)^2 \, ,
\end{split}
\end{equation}
where $w, x, y, z \in \mathbb{C}$ and
\begin{equation}
\begin{split}
W'(x) = g (x-a_1) (x-a_2) \, . \\[1.5ex]
\end{split}
\end{equation}
We take $g$, $a_1$ and $a_2$ as real parameters and set $a_1 > a_2$.
The geometry has singularities at $x = a_1$ and $x= a_2$, which are resolved by blowing up,
leaving 2-spheres at these points.
These 2-spheres are isolated minimal cycles inside the Calabi-Yau manifold.
A non-supersymmetric vacuum can be constructed
by wrapping $N_1$ D5-branes and $|N_2|$ anti-D5-branes over the 2-spheres at $x = a_1$ and $x= a_2$ respectively \cite{Aganagic:2006ex}.
In the present paper, we concentrate on the case with $N_1 = - N_2 =N$ (For the anti-branes, $N_2$ is negative, which will be clear below)
just for simplicity.
Generalizations to other cases are straightforward.
The vacuum can be metastable because the branes and anti-branes, wrapping the homologous but minimal cycles,
have to climb a potential barrier to annihilate each other.

\begin{figure}[!t]
  \begin{center}
  \vspace{-2cm}
          \includegraphics[clip, width=15cm]{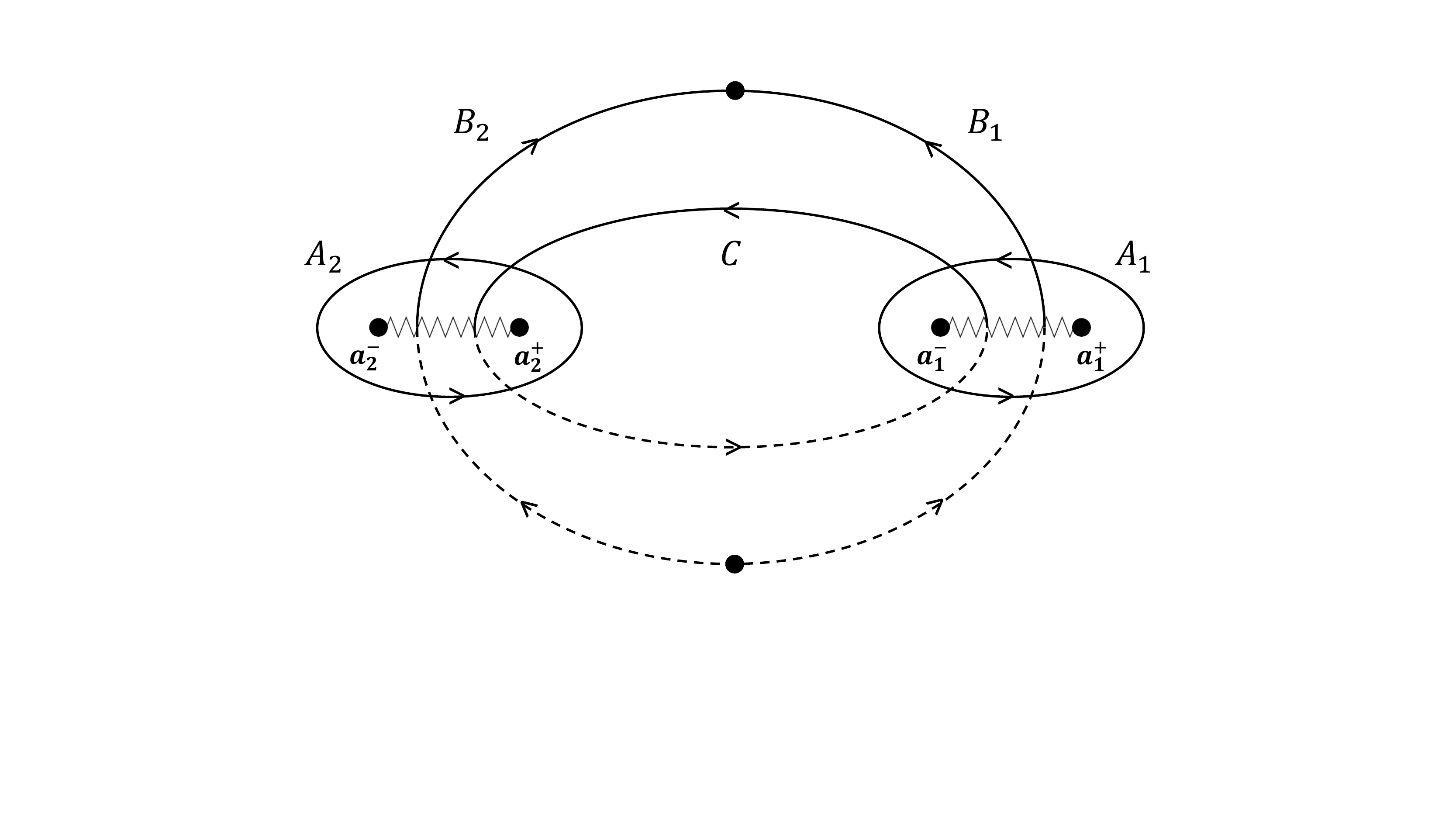}
  \vspace{-2cm}
    \caption{The geometry is projected to a double cover of the $x$ plane with $w = z = 0$.
The two sheets are laminated by two branch cuts.
The $A_i$ cycle corresponds to the contour around the branch cut
between $x = a_i^+$ and $x= a_i^-$ on the Riemann surface
while the $B_i$ cycle is represented by the contour connecting $x=a_i^+$ and a cutoff  of the noncompact cycle, $x=\Lambda_0$.
The $C (= B_{1} - B_2)$ cycle is denoted as the contour around $x=a_2^+$ and $x=a_{1}^-$.}
    \label{fig:riemann}
  \end{center}
\end{figure}

We now consider
the large $N$ dual of the above open string picture
\cite{Vafa:2000wi,Aganagic:2006ex,Cachazo:2001jy}.
By geometric transition, the 2-spheres are blown down
and the singularities at $x = a_1$ and $x= a_2$ are resolved into 3-spheres.
The branes and anti-branes wrapping the original 2-spheres are replaced with fluxes through the 3-spheres.
The defining equation of the geometry is changed into
\begin{equation}
\begin{split}
w z  = y^2 + W'(x)^2 + b_1 x + b_0\, , \label{geometry} \\[1.5ex]
\end{split}
\end{equation}
where $b_0$ and $b_1$ are parameters.
Each of the two critical points at $x = a_1$ and $x= a_2$ splits into two, $x = a_i^+$ and $x= a_i^-$ where $i = 1,2$.
The deformed geometry can be described by a Riemann surface, a double cover of the $x$ plane,
fibered over the coordinates $w$ and $z$.
The two sheets are laminated by two branch cuts.
The 3-cycle $A_i$ of the Calabi-Yau manifold is the above-mentioned 3-sphere and corresponds to the contour around the branch cut
between $x = a_i^+$ and $x= a_i^-$ on the Riemann surface
while the 3-cycle $B_i$ is represented by the contour connecting $x=a_i^+$ and a cutoff  of the noncompact cycle, $x=\Lambda_0$.
Furthermore, the $C$ cycle is a 3-cycle denoted as $C = B_{1} - B_2$
and represented by the contour around $x=a_2^+$ and $x=a_{1}^-$ on the Riemann surface.
See Figure~\ref{fig:riemann} for an illustration of these cycles on the Riemann surface.

In the closed string picture, there are nonzero fluxes through the $A_i$ and $B_i$ cycles,
\begin{equation}
\begin{split}
\\[-2.5ex]
\oint_{A_i} H \,=\, N_i \, , \qquad \int_{B_i} H \,=\, - \alpha \, , \qquad \oint_{C} \, H \,=\, 0 \, .
\label{fluxes}
\end{split}
\end{equation}
where
\begin{equation}
\begin{split}
 H = H ^{RR} + \tau H^{NS} \, . \\[1ex]
\end{split}
\end{equation}
Here, $H^{RR}$ denotes the RR 3-form flux, $H^{NS}$ the NSNS flux, and $\tau = C_0 + i / g_s$ is the type IIB dilaton-axion.
The flux number $N_i$ is positive or negative, depending on branes or anti-branes wrapping over the 2-sphere
in the original open string picture.
The flux through the $B_i$ cycle is given by the bare gauge coupling constant
defined at the cutoff scale $\Lambda_0$,
\begin{equation}
\begin{split}
\\[-2.5ex]
\alpha (\Lambda_0)  = - \frac{\theta_{\rm YM}}{2\pi} - i \frac{4\pi}{g_{\rm YM}^2 (\Lambda_0)} \, . \\[1ex]
\end{split}
\end{equation}
The periods of the 3-cycles are described in terms of the holomorphic $(3,0)$ form $\Omega$ as
\begin{equation}
\begin{split}
\\[-2.5ex]
\oint_{A_i} \Omega \,=\, S_i \, , \qquad  \int_{B_i} \Omega \,=\, \partial_{S_i} \mathcal{F}_0 \, , \qquad
\oint_{C} \, \Omega \,=\, \int_{B_1} \Omega - \int_{B_2} \Omega \, , \\[1.5ex]
\end{split}
\end{equation}
where $\mathcal{F}_0$ is the prepotential of the $\mathcal{N} = 2$ $U(1)$ gauge theory in type IIB string theory without fluxes.
The nonzero fluxes induce the superpotential \cite{Gukov:1999ya} as
\begin{equation}
\begin{split}
\\[-2.5ex]
\mathcal{W} (S) = \int H \wedge \Omega \,=\, \alpha (S_1 + S_2) + N \partial_{S_1} \mathcal{F}_0 - N \partial_{S_2} \mathcal{F}_0 \, .
\end{split}
\end{equation}
The period of the $B_i$ cycle was calculated in Ref.~\cite{Cachazo:2001jy} from which we obtain
the Kahler metric $g_{i \bar{j}} = {\rm Im} (\tau_{ij})$ where $\tau_{ij}$ is the period matrix of the Calabi-Yau,
$\tau_{ij} = \partial_{S_i} \partial_{S_j} \mathcal{F}_0$.
Then, we can compute the potential,
$V (S) = g^{i \bar{j}} \partial_{S_i} \mathcal{W} \, \overline{\partial_{S_j} \mathcal{W}}$.
Minimizing the potential, we find the periods of the $A_i$ and $C$ cycles
\cite{Aganagic:2006ex,Heckman:2007wk},
\begin{equation}
\begin{split}
\\[-2.5ex]
\oint_{A_1} \Omega \,\propto\, \zeta_1 g \Delta^3 e^{-2\pi i \alpha / N} \, , \qquad
\oint_{A_2} \Omega \,\propto\, - \zeta_2 g \Delta^3 e^{2\pi i \overline{\alpha} / N} \, , \qquad
\oint_{C} \, \Omega \,\sim\, g \Delta^3 \, , \\[1.5ex]
\end{split}
\end{equation}
where $\Delta = a_1 - a_2$ is the distance between the critical points before the transition.
The complex factor $\zeta_i$ denotes $N$ roots of unity for each cut
and indicates the existence of the multiple vacua in this system with the fixed number of the flux lines.
The periods of $A_1$ and $A_2$ cycles are of equal size and exponentially suppressed compared to that of the $C$ cycle.
If we take $\zeta_1 = \zeta_2 = 1$,
the parameter $\theta_{\rm YM}$ controls the relative phases of the periods of the $A_1$ and $A_2$ cycles.
Setting $\theta_{\rm YM} = 0$ means the brunch cuts are aligned along the real axis of the Riemann surface.
Changing  $\theta_{\rm YM}$, the cuts rotate in opposite direction \cite{Shigemori}.
In the following discussions, we concentrate on the case with $\theta_{\rm YM} = 0$.\footnote{
When we consider $\theta_{\rm YM}$ as a dynamical field,
higher order corrections generate a potential for $\theta_{\rm YM}$ which is stabilized at $\theta_{\rm YM} = 0$
\cite{Heckman:2007wk}.}

In the present geometry, there are many states classified in terms of the number of the flux lines through the $A$ cycles
or the directions of the branch cuts. 
The vacuum energy of these vacua at the leading order is given by
\cite{Aganagic:2006ex}
\begin{equation}
\begin{split}
\\[-2.5ex]
V_0 &\,=\, \frac{16 \pi }{g_{\rm YM}^2} N  - \frac{2}{\pi} N^2 \log \left| \frac{\Lambda_0}{\Delta} \right|^2  \, , 
\label{leadingenergy}\\[1.5ex]
\end{split}
\end{equation}
where the first and second terms correspond to the bare tension of the branes and the Coulomb attraction
between them respectively in the open string picture.
Higher order corrections,
calculated by an auxiliary matrix model
\cite{Dijkgraaf}, lift degeneracy of the different vacua for the directions of the branch cuts.
The splitting is calculated as
\cite{Heckman:2007wk}
\begin{equation}
\begin{split}
\\[-2.5ex]
\delta V_{\, l_1, l_2} &\,\propto\, - N^2 e^{-\frac{8\pi^2}{Ng_{\rm YM}^2}} \left( \cos \left( \frac{2\pi l_1}{N} \right) 
+ \cos \left( \frac{2\pi l_2}{N} \right)  \right) \, , \label{energycorrection} \\[1.5ex]
\end{split}
\end{equation}
where we have defined $\zeta_1 = \exp (2\pi i l_1 /N)$ and $\zeta_2 = \exp (2\pi i l_2 /N)$ for $l_1, l_2 = 0 \sim N-1$.
We can see that the vacuum where the brunch cuts align along the real axis of the Riemann surface,
$l_1 = l_2 = 0$, is energetically preferred.

As discussed in Ref.~\cite{Heckman:2007wk},
metastability of the vacua is lost when the Coulomb attraction contribution, the second term in \eqref{leadingenergy},
becomes comparable to the first bare tension term,
\begin{equation}
\begin{split}
\\[-2.5ex]
\frac{1}{g_{\rm YM}^2 N} &\,\sim\,  \frac{1}{8\pi^2} \log \left| \frac{\Lambda_0}{\Delta} \right|^2  \, . \label{metastability} \\[1.5ex]
\end{split}
\end{equation}
In this case, the branch cuts rotate toward the real axis of the $x$-plane and expand toward each other.
The flux lines through the $A$ cycles are annihilated.
In the present paper, we assume a suitable parameter set in \eqref{metastability} to retain metastability of the vacua.

\subsection{Phase transitions}

Phase transitions between the vacua with fluxes proceed
like brane/flux annihilation
\cite{Kachru:2002gs}.
There are two types of phase transitions in the present geometry,
changing the directions of the branch cuts and reducing the flux lines through the $A$ cycles.
Both proceed via wrapping a D5-brane over a 3-cycle in the Calabi-Yau manifold.
The decay rate can be estimated as $\Gamma \sim\exp (-S_{O(4)})$
in terms of the bounce action for an $O(4)$ symmetric bubble
\cite{Aganagic:2006ex},
\begin{equation}
\begin{split}
\\[-2.5ex]
S_{O(4)} \,=\, \frac{27 \pi^2}{2} \frac{T_{\rm DW}^4}{(\Delta V)^3} \, , \label{O(4)bounce} \\[1.5ex]
\end{split}
\end{equation}
where $\Delta V$ is the energy difference between vacua and
$T_{\rm DW}$ is the tension of the domain wall D5-brane.
As the domain wall tension is large or the energy difference is small,
the decay rate is suppressed.

\subsubsection{Transitions between the $N$ confining vacua}

We first consider phase transitions changing the directions of the branch cuts.
With a fixed number of the flux lines though the $A$ cycles,
\eqref{energycorrection} tells us that the lowest energy state is at $l_1 = l_2 =0$.
Let us look at the decay of the first exited state $l_i = N-1$ ($i = 1$ or $2$) to the ground state.\footnote{
The states with $l_i = 1$ ($i = 1$ or $2$)  are also the first excited ones.
We can consider the decays of these states in the same way.}
A D5-brane wrapping the $S^2$ part of the $A_i$ cycle
moves from one pole $x=a_i^-$ to the other pole $x=a_i^+$ of the cycle.
This process leads to
\begin{equation}
\begin{split}
\\[-2.5ex]
\int_{B_i (x= a_i^+)} H \, - \, \int_{B_i (x= a_i^-)} H   \,=\, 1  \, . \\[1.5ex]
\end{split}
\end{equation}
The ground state is reached,  $l_i \rightarrow l_i+1$.
Due to the tension of the D5-brane,
there is a potential barrier in this process and the excited state can be metastable.
The tension is written as
\begin{equation}
\begin{split}
T_{A_i} &\,=\, \frac{1}{g_s} \oint_{A_i} |\Omega| \, \equiv \, \frac{V_{A_i}}{g_s} \, , \\[1.5ex]
\end{split}
\end{equation}
where $V_{A_i}$ is the size of the $A_i$ cycle.
The energy difference between the vacua is given by
\begin{equation}
\begin{split}
\\[-2.5ex]
\Delta V(l_1 = N-1, l_2 = 0) &\,\propto\, N^2 e^{- \frac{8\pi^2}{N g_{\rm YM}^2}}\left( 1 - \cos \left( \frac{2\pi}{N} \right)   \right) \, . \\[1.5ex]
\end{split}
\end{equation}
Inserting them into the expression \eqref{O(4)bounce}, we can estimate the decay rate.

Next, for the decay of the second exited state $l_1 = l_2 = N-1$ to the ground state,
D5-branes wrap the $A_1$ and $A_2$ cycles.
The domain wall tension is given by the sum of $T_{A_1}$ and $T_{A_2}$,
and the energy difference is
\begin{equation}
\begin{split}
\Delta V(l_1 = N-1, l_2 = N-1) &\,\propto\, 2 N^2 e^{- \frac{8\pi^2}{N g_{\rm YM}^2}}\left( 1 - \cos \left( \frac{2\pi}{N} \right)   \right) \, . \\[1.5ex]
\end{split}
\end{equation}
For a higher energy state, the decay to a lower energy state typically proceeds by a single large fall in energy rather than a sequence of small falls via intermediate states.

\subsubsection{Transitions reducing the RR flux lines}

We here assume $l_1 = l_2 = 0$ and consider a transition to reduce the flux lines through the $A$ cycles.
The phase transition proceeds by wrapping a D5-brane over the $C$ cycle.
This process leads to one fewer number of the RR fluxes,
\begin{equation}
\begin{split}
\\[-2ex]
\oint_{A_1} H \,=\, N-1 \, , \qquad \oint_{A_2} H \,=\, - N+1 \, . \\[1.5ex]
\end{split}
\end{equation}
As above, there is a potential barrier in this process.
In the expression \eqref{O(4)bounce}, the tension of the domain wall is given by
\begin{equation}
\begin{split}
T_{C} &\,=\, \frac{1}{g_s} \, \oint_C |\Omega| \, \equiv \, \frac{V_C}{g_s} \, , \\[1.5ex]
\end{split}
\end{equation}
where $V_{C}$ is the size of the $C$ cycle.
The energy difference is obtained as
\begin{equation}
\begin{split}
\\[-2.5ex]
\Delta V (N \rightarrow N-1) \, \simeq \, \frac{ 16 \pi}{g_{\rm YM}^2}  \, . \\[1.5ex]
\end{split}
\end{equation}
For $N = \mathcal{O} (1)$, the open string picture becomes valid.
Then, brane/anti-brane annihilation proceeds inside the manifold.
In the general cases where the directions of the branch cuts are not aligned along the real axis,
Ref.~\cite{Heckman:2007wk} discussed that the transition reducing the RR flux lines proceeds most efficiently
with the branch cuts rotating toward the real axis of the Riemann surface.
In section $4$, we will concentrate on the second excited state in the $N$ confining vacua and discuss
the simultaneous phase transitions to rotate the directions of the brunch cuts and reduce the RR flux lines
with the catalytic effect from a baryon-like object.

\section{Dibaryon/monopole bound states}

We have presented the geometrical setup defined by \eqref{geometry} with fluxes \eqref{fluxes}
and reviewed the phase structure.
In this section, we propose a new type of baryonic objects in string theory.
Let us wrap a D3-brane over each $A$ cycle in the present geometry.
As discussed in Ref.~\cite{Witten:1998xy}, the presence of the RR fluxes induces
the charge of the gauge field on the D3-brane world-volume,
\begin{equation}
\begin{split}
&\int_{ {\bf R}\times A_1} H \wedge \mathcal{A}_1 \,=\, \oint_{ A_1 } H \int_{{\bf R}} \mathcal{A}_1
\,=\, N \int_{{\bf R}} \mathcal{A}_1\, , \\[3ex]
&\int_{ {\bf R}\times A_2} H \wedge \mathcal{A}_2 \,=\, \oint_{ A_2} H \int_{{\bf R}} \mathcal{A}_2
\,=\, -N \int_{{\bf R}} \mathcal{A}_2 \, , \\[2ex]
\end{split}
\end{equation}
where $\mathcal{A}_1$ and $\mathcal{A}_2$ are the gauge fields on the D3-branes wrapping the $A_1$ and $A_2$ cycles respectively.
${\bf R}$ denotes the time direction.
Note that the sign of the gauge charges is opposite for two cycles corresponding to the sign of the RR fluxes.

Since the total charge of a gauge field has to vanish in a closed cycle,
the D3-brane must be attached by fundamental strings, which provide other sources of the gauge charge,
and a baryon vertex is formed.
The vertices can be described in terms of antisymmetric tensors,
\begin{equation}
\begin{split}
\\[-2.5ex]
E_{p_1, \cdots , \, p_{N}} \, , \qquad \tilde{E}^{\, \tilde{p}_1, \cdots , \, \tilde{p}_{N}} \, , \\[1.5ex]
\end{split}
\end{equation}
for the D3-branes wrapping the $A_1$ and $A_2$ cycles respectively.
Here, the lower (upper) indices $p$ ($\tilde{p}$) run from $1$ to $N$.
Which branes are the opposite ends of the strings attached with?
One possibility is that  the D3-branes wrapping the $A$ cycles
can be connected with each other by the fundamental strings,
which is so-called dibaryon (see \cite{dibaryon} for example).
In the language of the effective field theory, this can be described as
\begin{equation}
\begin{split}
\\[-2.5ex]
E_{p_1, \cdots , \, p_{N}} \tilde{E}^{\, \tilde{p}_1, \cdots , \, \tilde{p}_{N}} Q^{p_1}_{\tilde{p}_1} \cdots \, Q^{p_N}_{\tilde{p}_N}   \, ,  \label{baryon} \\[1.5ex]
\end{split}
\end{equation}
where $Q$'s denote bi-fundamental quarks which come from the fundamental strings connecting the D3-branes.
In this case, the strings have to be stretched in the internal space with length between the $A_1$ and $A_2$ cycles.

There is a bound state of this operator with a monopole. Let us wrap a D3-brane over the $C$ cycle as well as the $A_1$ and $A_2$ cycles at the same point of our spacetime. In this case, the configuration where the D3-branes on the $A$ cycles are connected by the stretching strings is not energetically favorable. Dissolving the fundamental strings into the D3-brane on the $C$ cycle makes the total energy lower,
\begin{equation}
\begin{split}
\\[-2.5ex]
E_{p_1, \cdots , \, p_{N}} \tilde{E}^{\, \tilde{p}_1, \cdots , \, \tilde{p}_{N}} Q^{p_1}_{\tilde{p}_1} \cdots \, Q^{p_N}_{\tilde{p}_N}    M \, , \label{baryon2} \\[1.5ex]
\end{split}
\end{equation}
where $M$ represents the D3-brane wrapping the $C$ cycle.

The D3-brane wrapping the $C$ cycle can be identified with a magnetic monopole
as proposed by \cite{Kasai:2015exa} with the open string picture of the present geometry.
The previous section have shown that a phase transition reducing the RR flux lines proceeds
via wrapping a D5-brane over the $C$ cycle.
During the transition, it has been known that the D3-brane wrapping the same 3-cycle forms a bound state with the domain wall D5-brane.
In other words, the D3-brane is dissolved into the D5-brane with nontrivial magnetic fluxes of the brane gauge field.
Then, the D5/D3 bound state can be considered as a bubble with a magnetic monopole.
In the same way, we can discuss that the D3-branes wrapping the $A_1$ and $A_2$ cycles
are identified with magnetic objects as well as baryons.
During the phase transition between the $N$ confining vacua, D5-branes wrap the $A_1$ and $A_2$ cycles.
The D3-branes are dissolved into the D5-branes with nontrivial magnetic fluxes of the brane gauge fields.
In the next section, we will discuss the catalytic effect on a phase transition between vacua from the dibaryon.

Although concentrating on the case with $N_1 = - N_2 = N$ in the present paper,
we briefly comment on other cases with $N_1 \neq -N_2$ here.
In these cases, we can also wrap a D3-brane over each $A$ cycle and form baryon vertices.
However, the D3-branes wrapping the $A$ cycles
cannot be connected with each other by the fundamental strings
because the numbers of the strings needed for the two branes are different.
Then, the bound state like \eqref{baryon} is not formed.
On the other hand, when a D3-brane wraps the $C$ cycle, the total charge of the $C$ cycle
cannot be zero only with the fundamental strings
attached with the D3-branes wrapping $A_1$ and $A_2$ cycles.
We can satisfy this condition by introducing nonzero background RR fluxes on the $C$ cycle.
In this case, generalizations of the bound state \eqref{baryon2} can be easily constructed.
We leave the analysis of the catalytic effects in the general cases to a future study.

\section{The catalytic effects}

We now study catalytic effects for the phase transitions discussed in section $2$ from the proposed dibaryons as impurities.
We have reviewed there are two kinds of transitions in the present geometry: the transition between the $N$ confining vacua and the one reducing the flux lines through the $A$ cycles.
They proceed by  wrapping domain wall D5-branes over the $A_{1,2}$ cycle and the $C$ cycle respectively.
As discussed above, in the presence of a dibaryon,
a bound state with a domain wall is formed
and the magnetic flux is induced on the wall.
We will analyze the possible catalytic effect on the various phase transitions one by one.

Let us first consider the decay of the ground state in the $N$ confining vacua
to a vacuum with a smaller number of the flux lines through the $A$ cycles.
This decay is similar to the one considered in Ref.~\cite{Kasai:2015exa}
where the magnetic monopole accelerates the decay of a false vacuum.
For the dibaryon like \eqref{baryon} in the present closed string picture,
the D3-branes wrap the $A_1$ and $A_2$ cycles so that
there is no catalytic effect on this decay.
On the other hand, the dibaryon like \eqref{baryon2} in fact accelerates the decay
because the D3-brane wraps the $C$ cycle.
For the calculation of the decay rate, see Ref.~\cite{Kasai:2015exa}.
Next, consider the catalytic effect on the transition between the $N$ confining vacua.
In this case, both the bound states of \eqref{baryon} and \eqref{baryon2} accelerate the decay.
Although the D3-brane wraps the different 3-cycle,
the calculation of the decay rate is done exactly in the same way with that in Ref.~\cite{Kasai:2015exa}.

Our main focus is the simultaneous phase transitions to rotate the directions of the cuts and reduce the number of the RR flux lines.
When we start with an excited state in  the $N$ confining vacua, this kind of transitions in fact occurs because
the flux lines through the $A$ cycles annihilate most efficiently with the branch cuts rotating toward the real axis of the Riemann surface.
If the theory is in the first excited state  $l_i = N-1$ ($i = 1$ or $2$),
the decay to the lower energy state with the cuts aligned along the real axis and a fewer number of the RR flux lines
proceeds by wrapping a D5-brane over both the $A_i$ cycle and the $C$ cycle.
In this case, the D3-branes wrapping the $A_i$ cycle and the $C$ cycle contribute to the decay rate.
Instead of analyzing this case, we will concentrate on the most interesting case where the catalytic effect is maximized,
the decay of the second excited state $l_1 = l_2 = N-1$ below.
Then, we can expect that all the D3-branes in the bound state of the dibaryon \eqref{baryon2} contribute to the decay rate.
As we will see, the potential energy of this system can have multiple local minima and the estimation of the decay rate is
not a trivial generalization of the above cases.
We discuss the dibaryon can in fact accelerate the decay of the metastable vacuum
and sometimes threaten the stability itself.

\subsection{Metastability of the vacuum with impurity}

We first investigate dependence of metastability of a false vacuum on the strength of the magnetic flux.
We concentrate on the decay of the second exited state $l_1 = l_2 = N-1$ to the state $l_1 = l_2 = 0$
with one fewer number of the RR fluxes through the $A$ cycles.
Since we assume $N_1 = - N_2 = N$, the size of the $A_1$ and $A_2$ cycles is the same,
\begin{equation}
\begin{split}
\\[-2.5ex]
{1\over 2} a \,\equiv\, {V_{A_1}\over V_C  } \,=\, {V_{A_2}\over V_C  } \, , \\[1.5ex]
\end{split}
\end{equation}
where $a$ is much smaller than one.
We consider the $C$ cycle wrapped by $n_{D3}$ D3-branes and each $A_i$ cycle wrapped by $\tilde{n}_{D3}$ D3-branes.
These D3-branes form a bound state with a domain wall D5-brane wrapping the $A$ and $C$ cycles
during the phase transition.
The total energy of the bubble with a radius $R$ in our spacetime is given by
\begin{equation}
\begin{split}
\\[-2ex]
E_{\, \rm total } &= \sqrt{\left(  4 \pi R ^2 T_{D5} V_C \right)^2 + \left( n_{D3} T_{D3} V_C \right)^2}
+ a \sqrt{\left( 4 \pi R ^2 T_{D5}  V_C \right)^2 + \left( \tilde{n}_{D3} T_{D3} V_C \right)^2} \\[1.5ex]
&\quad  -{4\pi \over 3} R^3 \Delta V \, , \label{totalenergy} \\[1.5ex]
\end{split}
\end{equation}
where $T_{D5}$ and $T_{D3}$ are the tensions of a D5-brane and a D3-brane respectively.
The first and second terms represent the energy of the bound state consisting of the dibaryon and the domain wall D5-brane.
The third term is the benefit of energy inside the bubble.
$\Delta V$ is the energy difference between the false vacuum and the lower state.

\begin{figure}[t]
\begin{center}
\vspace{-0.5cm}
\includegraphics[clip, width=6cm]{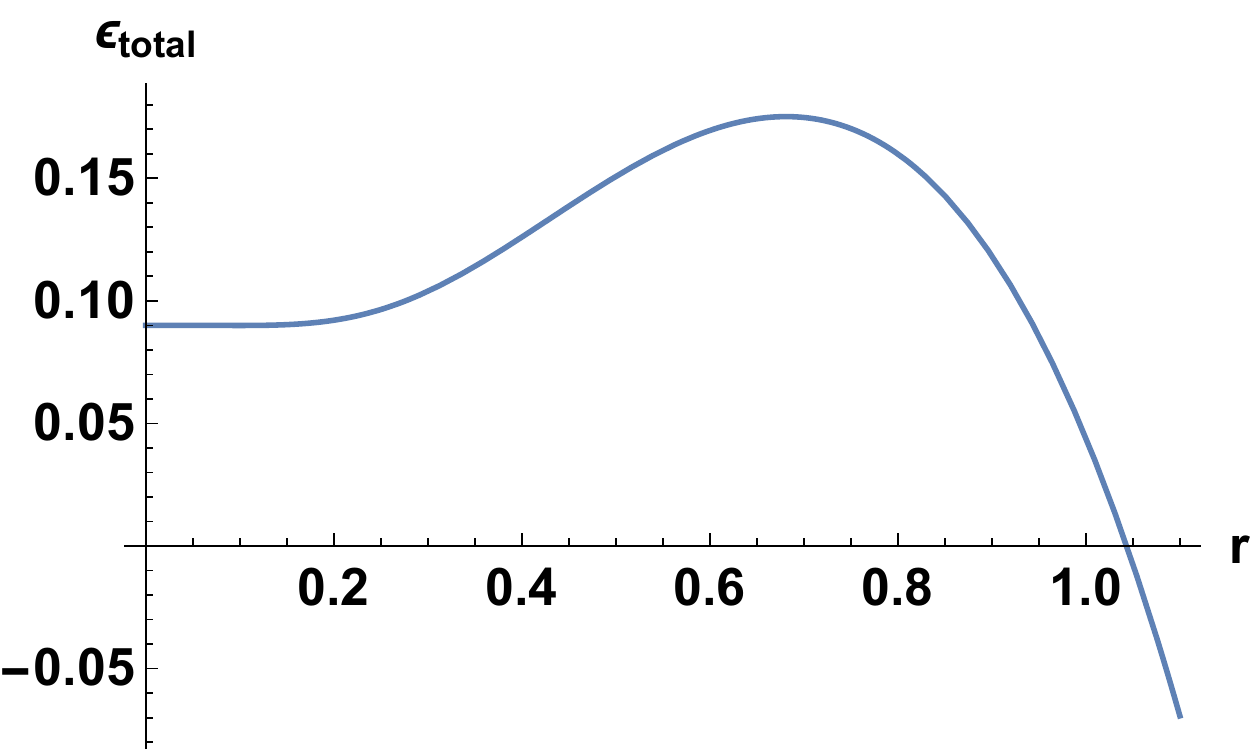} \hspace{2cm}
\includegraphics[clip, width=6cm]{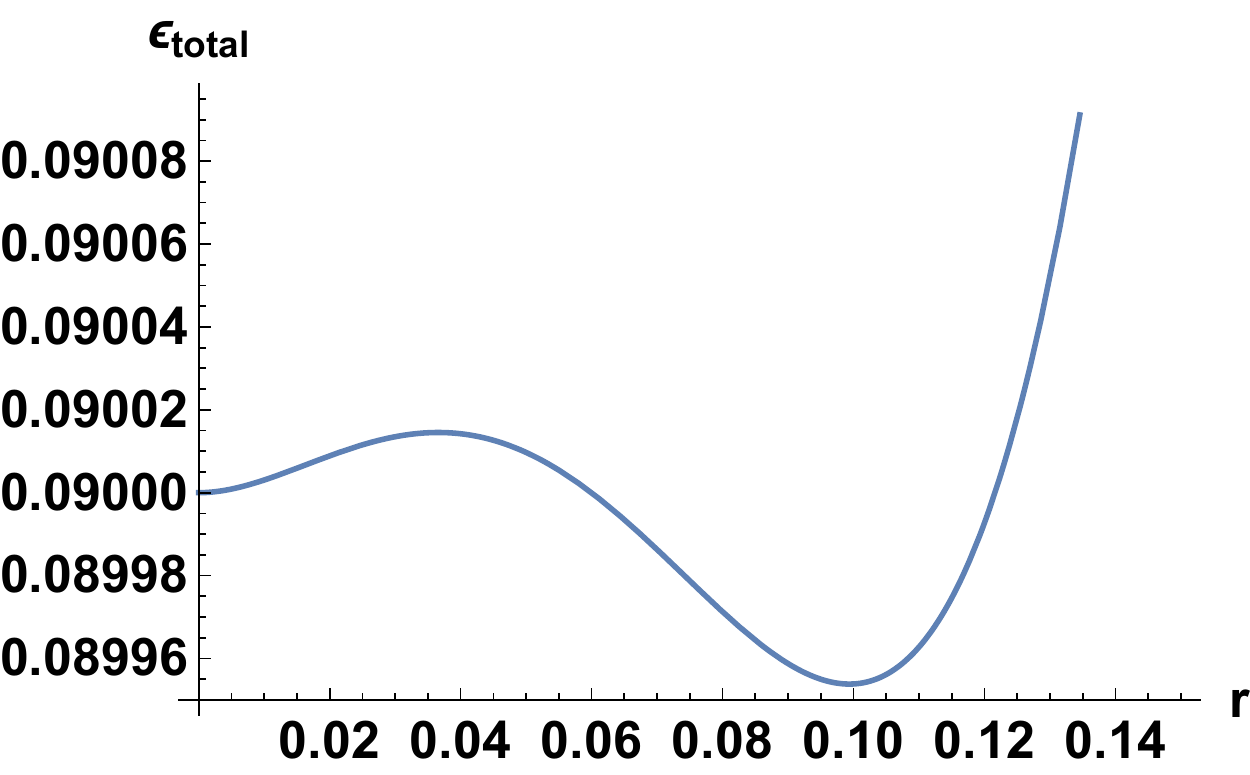} \vspace{0.5cm}
\caption{The dimensionless energy $\epsilon_{\rm total}$
as a function of the dimensionless radius of the domain wall $r$ for $a=0.04$, $b= 0.3$, $\tilde{b} = 0$.
The right panel shows the enlarged view of the left one near the origin of $r$.
The local minima are at $r=0$ and $r\simeq 0.1$.}
\label{fig:Energy1}
\end{center}
\end{figure}
\begin{figure}[!t]
\begin{center}
\includegraphics[clip, width=6cm]{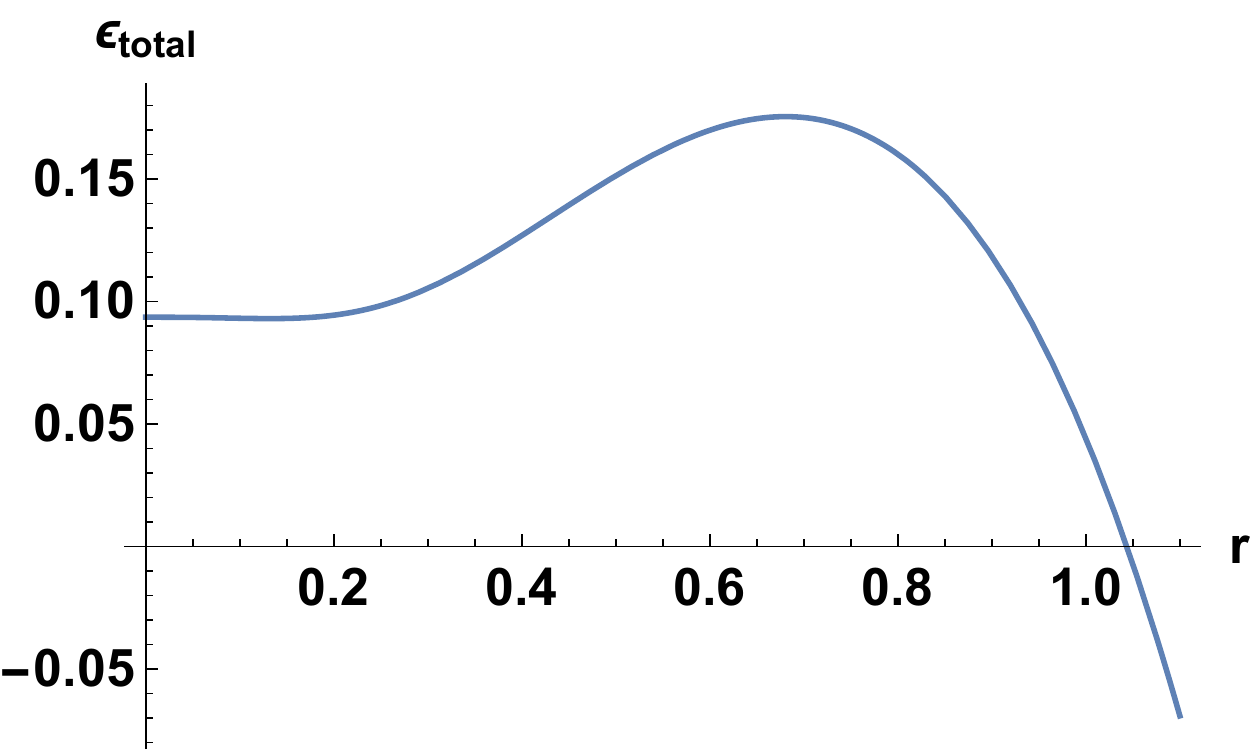} \hspace{2cm}
\includegraphics[clip, width=6cm]{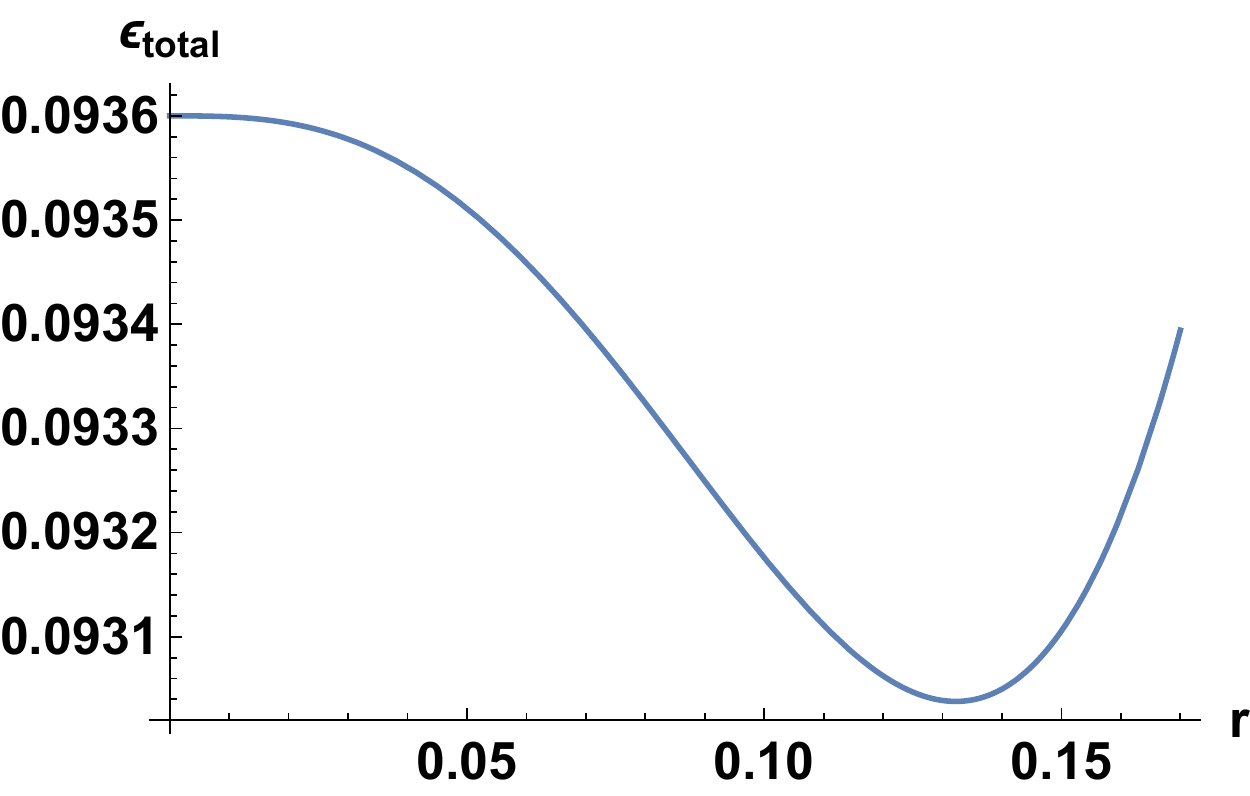} \vspace{0.5cm}
\caption{The dimensionless energy $\epsilon_{\rm total}$ in the case with $a=0.04$, $b= 0.3$, $\tilde{b} = 0.3$.
The right panel is the enlarged view.
The local minimum is at $r\simeq 0.1$.}
\label{fig:Energy2}
\end{center}
\end{figure}

\begin{figure}[!t]
\begin{center}
\includegraphics[clip, width=6cm]{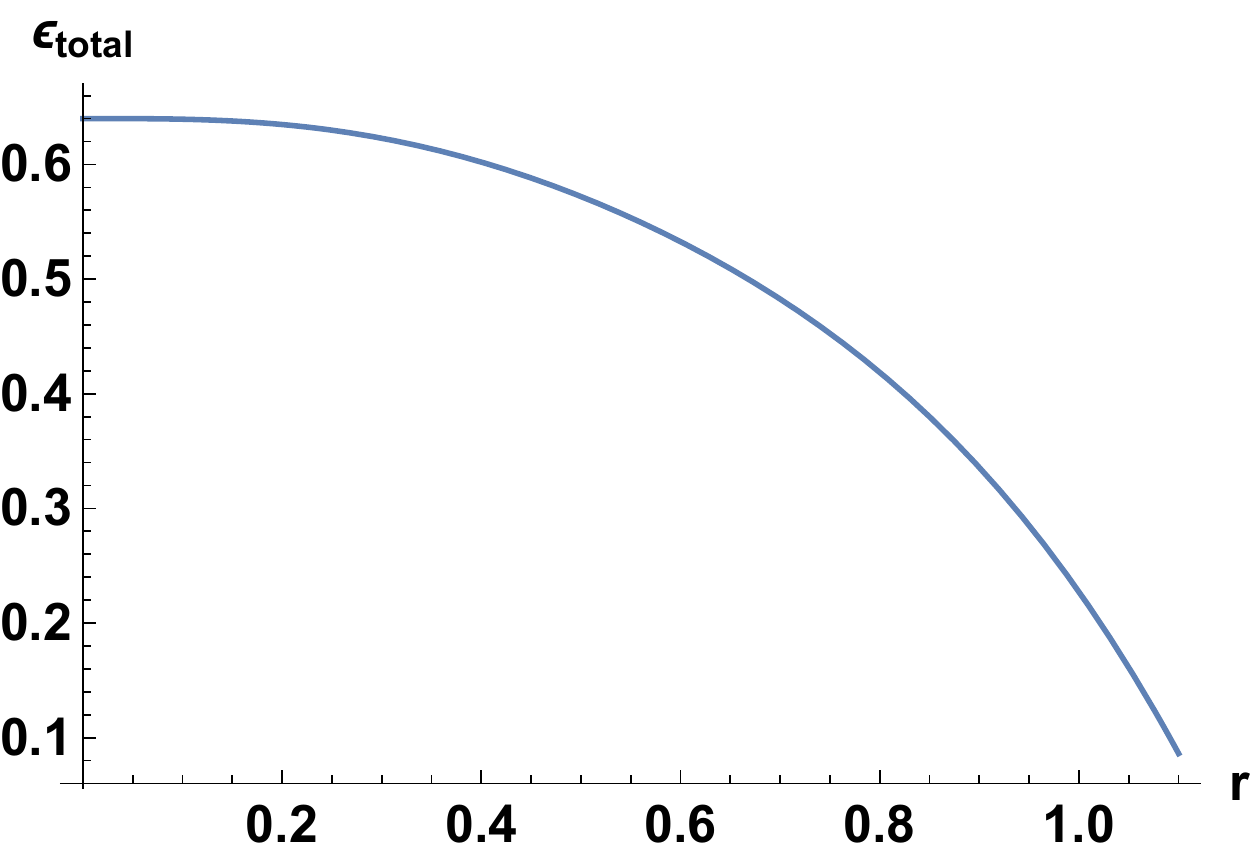} \hspace{2cm}
\includegraphics[clip, width=6cm]{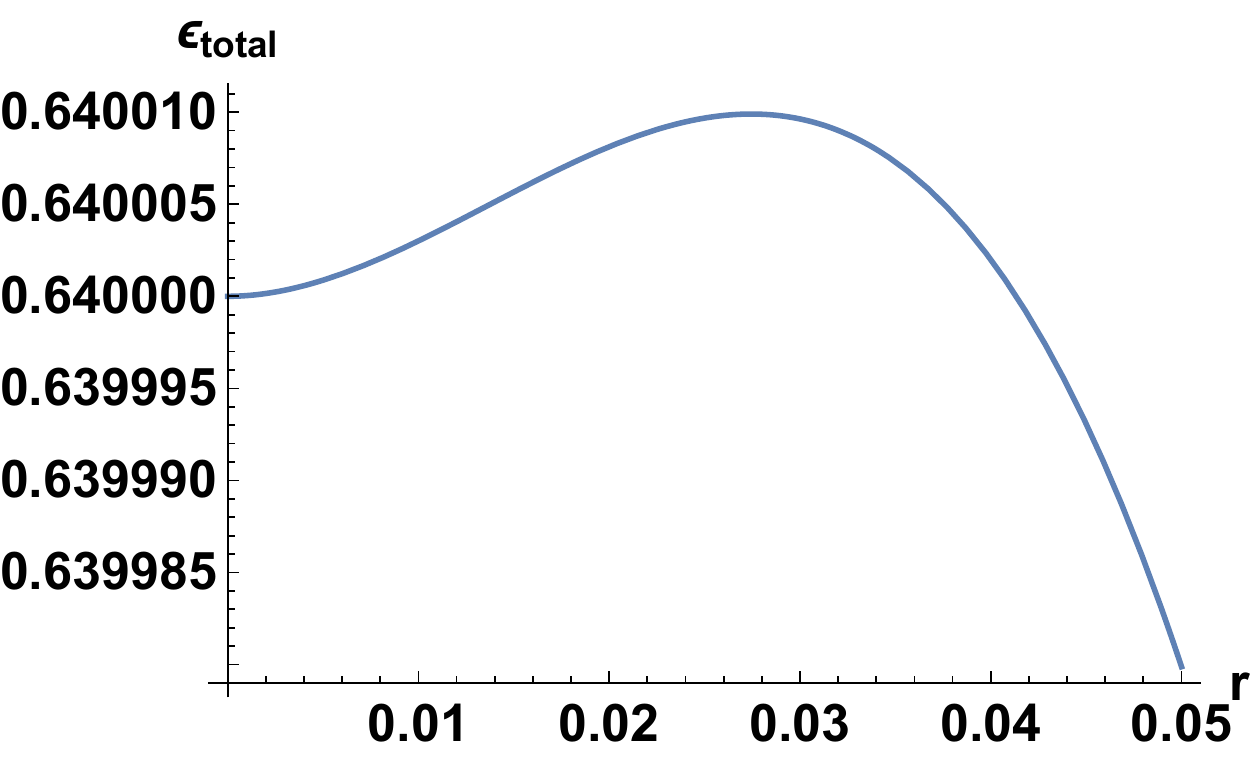} \vspace{0.5cm}
\caption{The dimensionless energy $\epsilon_{\rm total}$ in the case with $a=0.04$, $b= 0.8$, $\tilde{b} = 0$.
The right panel is the enlarged view.
The local minimum is at $r=0$.}
\label{fig:Energy3}
\end{center}
\end{figure}

\begin{figure}[!t]
\begin{center}
\includegraphics[clip, width=6cm]{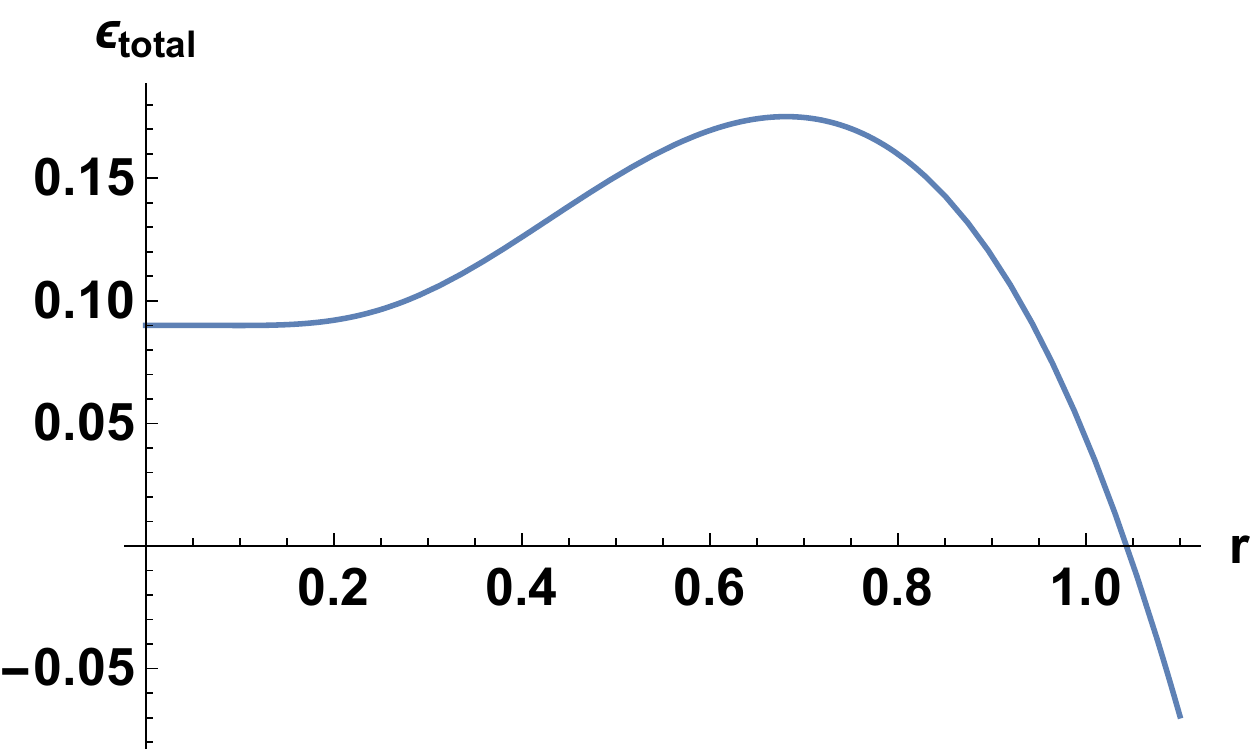} \hspace{2cm}
\includegraphics[clip, width=6cm]{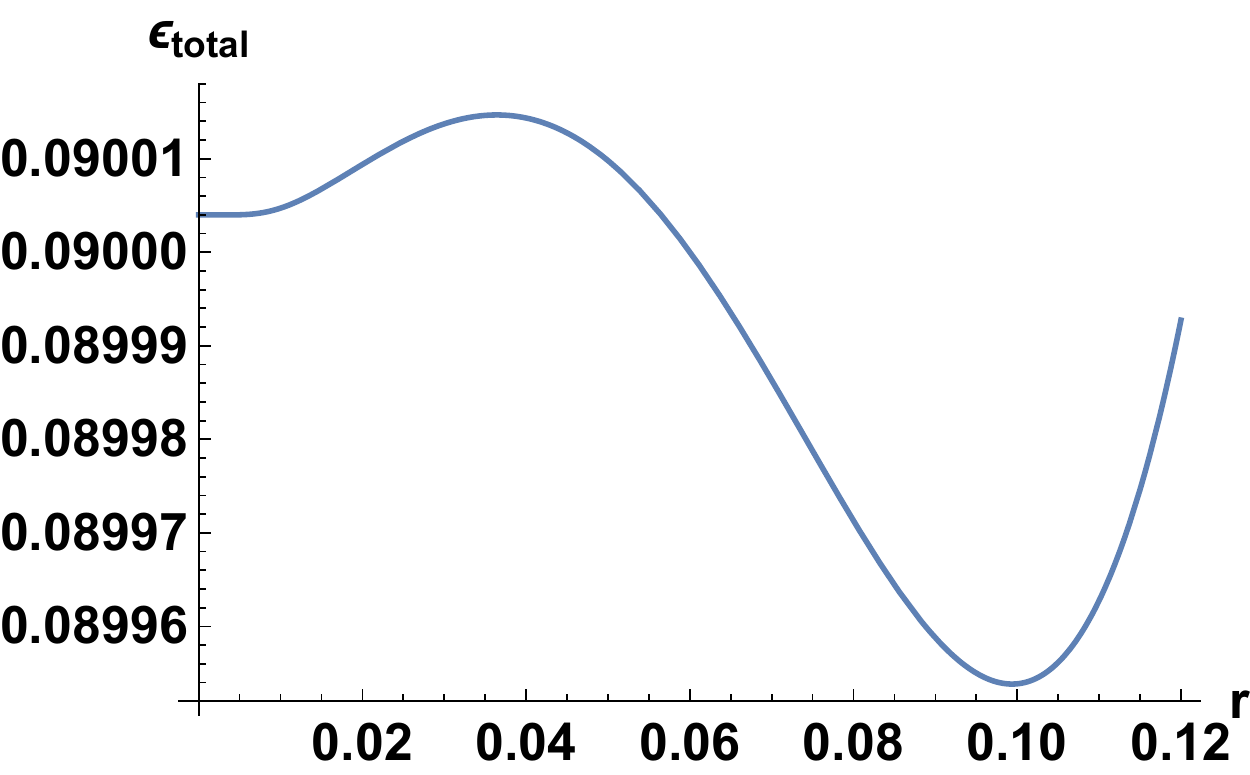} \vspace{0.5cm}
\caption{The dimensionless energy $\epsilon_{\rm total}$ in the case with $a=0.04$, $b= 0.3$, $\tilde{b} = 0.01$.
The right panel is the enlarged view.
The local minima are at $r\simeq0.004$ and $r\simeq 0.1$.}
\label{fig:Energy4}
\end{center}
\end{figure}

Let us look at the shape of the total energy \eqref{totalenergy} in more detail.
In order to do so, we now define the following dimensionless quantities:
\begin{equation}
\begin{split}
\\[-2.5ex]
r \equiv \frac{\Delta V}{3 T_{D5} V_C} R  \, , \qquad b^4 \equiv \left( \frac{\Delta V}{3 T_{D5} V_C} \right)^4
\left( \frac{n_{D3} T_{D3}}{4 \pi T_{D5}} \right)^2
\, , \qquad \tilde{b}^4 \equiv \left( \frac{\tilde{n}_{D3}}{n_{D3}} \right)^2 b^4  \, , \label{dimensionless} \\[1.5ex]
\end{split}
\end{equation}
where the parameters $r$, $b$, $\tilde{b}$ represent the radius of the bubble in the phase transition,
the magnetic flux induced by the D3- branes wrapping the $C$ cycle and the flux by the D3-branes wrapping the $A$ cycles respectively.
In terms of these dimensionless quantities, we can rewrite the expression of the total energy \eqref{totalenergy} as
\begin{equation}
\begin{split}
\\[-2.5ex]
E_{\, \rm total } = \frac{36 \pi (T_{D5} V_C)^3}{(\Delta V)^2}\left( \sqrt{r^4+b^4}+a\sqrt{r^4+\tilde{b}^4}-r^3   \right) 
\equiv \frac{36 \pi (T_{D5} V_C)^3}{(\Delta V)^2} \, \epsilon_{\rm total} \, , \\[1.5ex]
\end{split}
\end{equation}
where we have defined the dimensionless total energy $\epsilon_{\rm total}$.
We plot $\epsilon_{\rm total}$ in Figure~\ref{fig:Energy1}, \ref{fig:Energy2}, \ref{fig:Energy3}, \ref{fig:Energy4}
as a function of the dimensionless radius of the domain wall $r$
for different values of $b$ and $\tilde{b}$.
The right panels of these figures show the enlarged views of the left panels near the origin of $r$.
We fix the ratio of the size of the $A$ and $C$ cycles as $a = 0.04$.

In Figure~\ref{fig:Energy1}, we take $b=0.3$ and turn off $\tilde{b}$, that is, wrap no D3-branes over the $A$ cycles.
This setup is similar to the case considered in
\cite{Kasai:2015exa} and the impurity is in fact a magnetic monopole while there are finite sized $A$ cycles in the present case.
We can see that there are two local minima at $r=0$ and $r\simeq 0.1$ and potential barriers.
In this case, the phase transition to the lower energy state proceeds via two sequential quantum tunneling.
In Figure~\ref{fig:Energy2}, we wrap D3-branes over the $A$ cycles, $\tilde{b} = 0.3$, as well as taking $b = 0.3$.
In this case, there is no local minimum at $r=0$ due to
the catalytic effect from the dibaryon.
Figure~\ref{fig:Energy3} shows the case with $b=0.8$ and $\tilde{b} =0$.
We see that there is only one local minimum at $r=0$,
which means the catalytic effect from the magnetic monopole vanishes one potential barrier present in the above two cases.
The phase transition proceeds via just one quantum tunneling.
In Figures~\ref{fig:Energy4}, the case with $b = 0.3$ and $\tilde{b} = 0.01$ is shown.
We see that there are two local minima both at nonzero values of $r$ and potential barriers.
The phase transition proceeds via two sequential tunneling, as we will see below.


\begin{figure}[!t]
\begin{center}
\vspace{-0.5cm}
\includegraphics[clip, width=6.5cm]{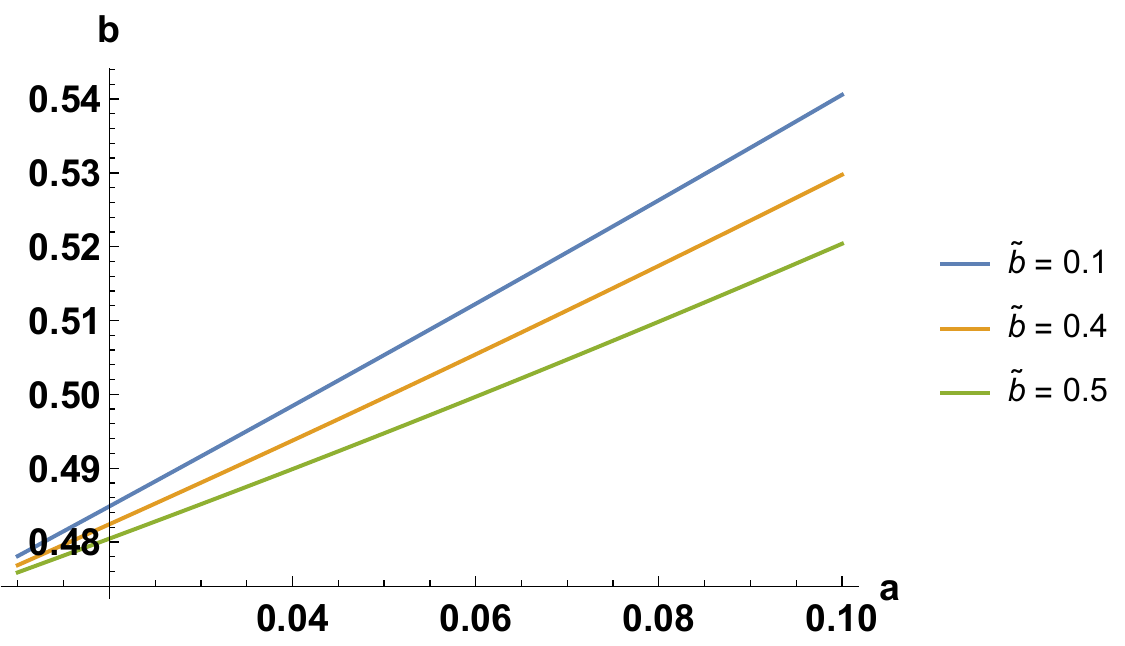} \hspace{2cm}
\includegraphics[clip, width=6.5cm]{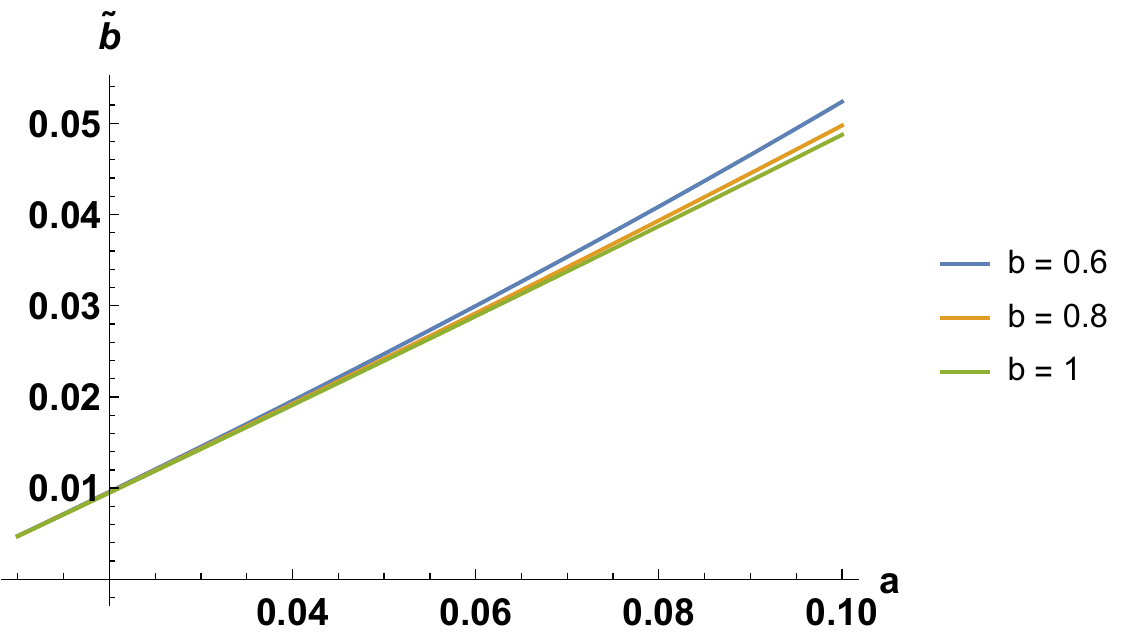} \vspace{0.5cm}
\caption{The conditions for $b$ (left) and $\tilde{b}$ (right) for the false vacuum to be unstable, changing a value of $a$.
In the region above the lines, the vacuum is unstable.}
\label{fig:unstable}
\end{center}
\end{figure}

We then consider the condition for the (non)existence of the false vacuum.
We see this by searching parameter sets of $a$, $b$ and $\tilde{b}$ where there is no local minimum in
the dimensionless energy $\epsilon_{\rm total}$ for any value of $r$.
This can be expressed as
\begin{equation}
\begin{split}
\\[-2.5ex]
\frac{d \epsilon_{\rm total} (r)}{dr} \leq 0 \qquad \text{for any $r$} \, . \label{unstablecondition} \\[1.5ex]
\end{split}
\end{equation}
We can rewrite this condition to the one for $b$ as follows:
\begin{equation}
\begin{split}
\\[-2.5ex]
b^4\,\ge\, 4r^6 \left( {-2ar^3\over \sqrt{r^4+{\tilde{b}}^4}} +3r^2 \right)^{-2}-r^4 \qquad \text{for any $r$} \, . \\[1.5ex]
\end{split}
\end{equation}
When there is no local minimum in $\epsilon_{\rm total}$,
the maximum value of the right hand side is smaller than the left hand side.
When $\tilde{b} < \frac{\sqrt{2}}{3}a$ is satisfied, the parenthesis in the above inequality can be always zero at some value of $r$
so that the inequality cannot be satisfied, corresponding to the appearance of a local minimum.
On the other hand, we can also rewrite the condition \eqref{unstablecondition} to the one for $\tilde{b}$,
\begin{equation}
\begin{split}
\tilde{b}^4\,\ge\, 4a^2 r^6 \left( {-2r^3\over \sqrt{r^4+{{b}}^4}} +3r^2 \right)^{-2}-r^4 \, . \\[1.5ex]
\end{split}
\end{equation}
As above, when there is no local minimum in $\epsilon_{\rm total}$,
the maximum value of the right hand side is smaller than $\tilde{b}^4$.
When ${b} < \frac{\sqrt{2}}{3}$ is satisfied,
the inequality cannot be satisfied, suggesting the appearance of a local minimum.
Figure~\ref{fig:unstable} shows these conditions for $b$ (left) and $\tilde{b}$ (right), changing a value of $a$.
We plot the cases with $\tilde{b} = 0.1, 0.4, 0.5$ in the left panel while the cases with ${b} = 0.6, 0.8, 1$ in the right panel.
In the region above the lines, the vacuum is unstable.
Below, we avoid the parameter sets of $a$, $b$ and $\tilde{b}$ to satisfy these conditions and
assume that there is at least one local minimum in $\epsilon_{\rm total}$.

\subsection{The decay rate}

We now estimate the decay rate of the false vacuum, basically along the lines of \cite{Hashi,hyaku,PTK}.
The transition proceeds via a domain wall D5-brane wrapping over the $A_1$, $A_2$ and $C$ cycles
at the same point of our spacetime.
The decay rate is calculated by using the Euclidean version of the Dirac-Born-Infeld (DBI) action.
The coordinates of the $O(3)$ symmetric bubble is given by
\begin{equation}
\begin{split}
\\[-2.5ex]
X^0 = t \, , \quad X^1 = R(t) \sin \theta \cos \varphi \, , \quad X^2 = R(t) \sin \theta \sin \varphi
\, , \quad X^3 = R(t) \cos \theta   \, , \\[1.5ex]
\end{split}
\end{equation}
while $X^{I}$ ($I = 4,5,6$) are filling the $A_1$, $A_2$ and $C$ cycles.
During the phase transition, the baryon forms a bound state with the domain wall D5-brane
and induces the magnetic flux on the wall.
The magnetic flux  for the $C$ cycle is quantized as $\Phi \equiv 4 \pi R^2 {B'} = - 2 \pi n_{\rm D3}$
while the flux  for the $A_1$ and $A_2$ cycles is as $\widetilde{\Phi} \equiv 4 \pi R^2 \widetilde{B}' = - 2 \pi \tilde{n}_{\rm D3}$.
Here, ${B'}$ and $ \widetilde{B}'$ are defined as
$ R^2 \sin \theta \cdot {B'} \equiv  F_{\theta \varphi}$ and 
$ R^2 \sin \theta \cdot  \widetilde{B}' \equiv  \widetilde{F}_{\theta \varphi}$ respectively.
We now define $B \equiv 2 \pi \alpha' R^2 {B'}$ and $\widetilde{B} \equiv 2 \pi \alpha' R^2 \widetilde{B}'$
(We write the dependence of $\alpha'$ explicitly in this subsection).
Then, we find
\begin{equation}
\begin{split}
\\[-2.5ex]
- \det \left( \partial_\alpha X^\mu \partial_\beta X_\mu + 2 \pi \alpha' F_{\alpha \beta} \right)
&= - \det \left( \partial_a X^\mu \partial_b X_\mu + 2 \pi \alpha' F_{a b} \right)
\cdot \det \left( \partial_A X^I \partial_B X_I  \right)   \\[1.5ex]
&= \sin^2 \theta \, ( 1 - \dot{R}^2 ) (R^4 + B^2) \cdot \det \left( \partial_A X^I \partial_B X_I  \right) \, , \\[1.5ex]
\end{split}
\end{equation}
for the $C$ cycle and
\begin{equation}
\begin{split}
\\[-2.5ex]
- \det \left( \partial_\alpha X^\mu \partial_\beta X_\mu + 2 \pi \alpha' \widetilde{F}_{\alpha \beta} \right)
&= - \det \left( \partial_a X^\mu \partial_b X_\mu + 2 \pi \alpha' \widetilde{F}_{a b} \right)
\cdot \det \left( \partial_A X^I \partial_B X_I  \right)   \\[1.5ex]
&= \sin^2 \theta \, ( 1 - \dot{R}^2 ) (R^4 + \widetilde{B}^2) \cdot \det \left( \partial_A X^I \partial_B X_I  \right) \, , \\[1.5ex]
\end{split}
\end{equation}
for the $A_1$ and $A_2$ cycles.
Here, $a,b = (t, \theta , \varphi)$,
$A, B$ denote the coordinates of the $A_1$, $A_2$ and $C$ cycles and
$\alpha = (a, A)$, $\beta = (b, B)$.
The dot here represents the time derivative.
Since there are no off-diagonal parts of the matrices, the determinants are given by the products of two pieces.
The DBI action is then given by
\begin{equation}
\begin{split}
\\[-2ex]
S \,=\,   -\, T_{\rm D5} &\int_{C} d^6 \xi \, \sqrt{ \det \left( \partial_A X^I \partial_B X_I  \right) \strut}
\sqrt{- \det \left( \partial_a X^\mu \partial_b X_\mu + 2 \pi \alpha' F_{a b} \right) \strut} \\[1.5ex]
 -\, T_{\rm D5} &\int_{A_1+A_2} d^6 \xi \, \sqrt{ \det \left( \partial_A X^I \partial_B X_I  \right) \strut}
\sqrt{- \det \left( \partial_a X^\mu \partial_b X_\mu + 2 \pi \alpha' \widetilde{F}_{a b} \right) \strut} \\[1.5ex]
+ &\int dt \, \frac{4 \pi}{3} R^3 \Delta V  \nonumber \\[2ex]
\end{split}
\end{equation}
\begin{equation}
\begin{split}
\,=\,-\, T_{D5} &\int dt  \, \, 4\pi  V_C \, \sqrt{1-\dot{R}^2 \strut}
\, \left( \sqrt{R^4 +B^2 \strut} + a \sqrt{R^4 + \widetilde{B}^2} \right) \\[1.5ex]
+ &\int dt \, \frac{4 \pi}{3} R^3 \Delta V \, , \\[1.5ex]
\end{split}
\end{equation}
where we have used the fact that the integration with the indices $A, B$ is
given by the volume of the $C$ cycle or the $A_{1, 2}$ cycle.
We consider the Euclidean action $S_E$ by wick rotation, $t \rightarrow i \tau$, to calculate the decay rate.
We also define the following variable,
\begin{equation}
\begin{split}
\\[-2.5ex]
s \,\equiv\, \frac{\Delta V}{3 T_{D5}V_C} \tau \, . \\[1.5ex]
\end{split}
\end{equation}
Then, the Euclidean action is given by
\begin{equation}
\begin{split}
\\[-1ex]
S_E &\,=\, \left( \frac{27 \pi^2 }{2} \frac{( T_{D5} V_C)^4}{(\Delta V)^3} \right) \frac{8}{\pi}
\int ds \left( \sqrt{1+\dot{r}^2 \strut} \, F(r) -r^3 \right)
\,\equiv\, S_{O(4)} \, \frac{8}{\pi} \, S_{O(3)} \, , \label{Euclideanaction}\\[2.5ex]
\end{split}
\end{equation}
where we have used \eqref{O(4)bounce} with the identification $T_{\rm DW} = T_{D5} V_C$ and defined
\begin{equation}
\begin{split}
\\[-2.5ex]
F(r) = \sqrt{r^4+b^4 \strut}+a\sqrt{r^4+\tilde{b}^4 \strut} \, . \\[1.5ex]
\end{split}
\end{equation}
In addition, we have used $T_{D3}/ T_{D5} = 4 \pi^2 \alpha'$ and the dimensionless parameters \eqref{dimensionless}.
The dot in $\dot{r}$ of the action \eqref{Euclideanaction} denotes the derivative with respect to $s$.
The exponent of the decay rate is given by the difference between the on-shell Euclidean action and the integral of
the dimensionless energy $\epsilon_{\rm total} (r_{\rm min})$ at a local minimum,
\begin{equation}
\begin{split}
\\[-2.5ex]
S_{O(4)} \, \frac{8}{\pi} \, \Delta S_{O(3)} \,\equiv\, S_{O(4)} \, \frac{8}{\pi} 
\left( S_{O(3)}^{\, \rm on-shell} - \int ds \, \epsilon_{\rm total} (r_{\rm min}) \right) \, . \\[1.5ex]
\end{split}
\end{equation}
The equation of motion derived from the Euclidean action \eqref{Euclideanaction} is represented by the first order differential equation,
\begin{equation}
\begin{split}
\frac{d}{ds} \left( \frac{F(r)}{\sqrt{1+\dot{r}^2}} -r^3 \right) = 0 \, . \\[1.5ex]
\end{split}
\end{equation}
Then, we can obtain the velocity $\dot{r}$ in terms of $r$ and an integration constant $C$ as
\begin{equation}
\begin{split}
\\[-2.5ex]
\dot{r} = \pm \frac{1}{C+r^3} \sqrt{F(r)^2 - (C+r^3)^2 \strut} \, . \label{velocity}  \\[1.5ex]
\end{split}
\end{equation}
The integration constant is evaluated from the fact that the velocity is zero at the start point of the decay.
This is then given by
\begin{equation}
\begin{split}
\\[-2.5ex]
C = F(r_{\rm min}) - r_{\rm min}^3 \, . \\[1.5ex]
\end{split}
\end{equation}
Finally, we obtain
\begin{equation}
\begin{split}
\\[-2.5ex]
\Delta S_{O(3) } \,=\, \int^{r_{\rm max}}_{r_{\rm min}} dr \, \frac{1}{\dot{r}}  \left[ \, \left( \sqrt{1+\dot{r}^2 \strut} \, F(r) -r^3 \right)
- \epsilon_{\rm total} (r_{\rm min}) \right] \, , \\[1.5ex]
\end{split}
\end{equation}
where the integration is done until $r = r_{\rm max}$ where the velocity \eqref{velocity} vanishes other than the start point.

\begin{figure}[t]
\begin{center}
\vspace{-0.5cm}
\includegraphics[clip, width=6cm]{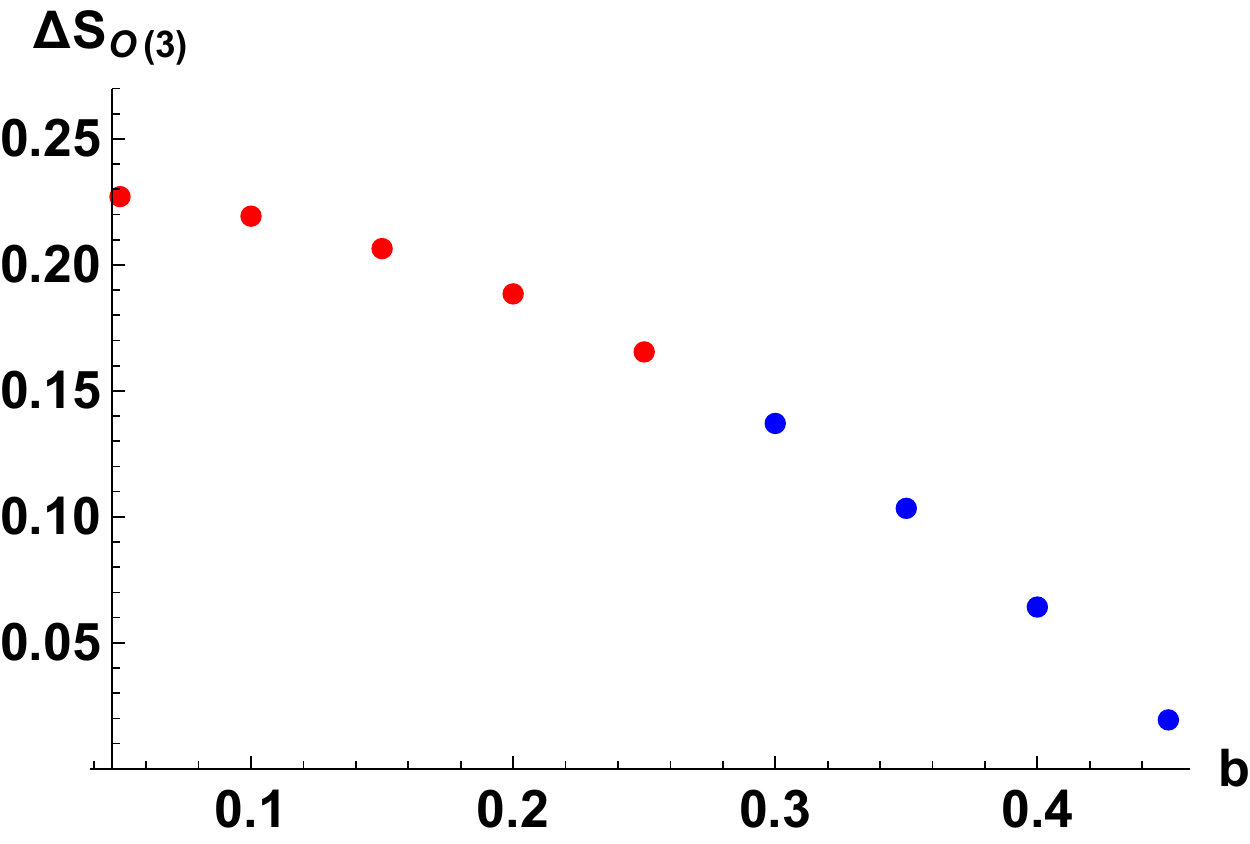} \hspace{2cm}
\includegraphics[clip, width=6cm]{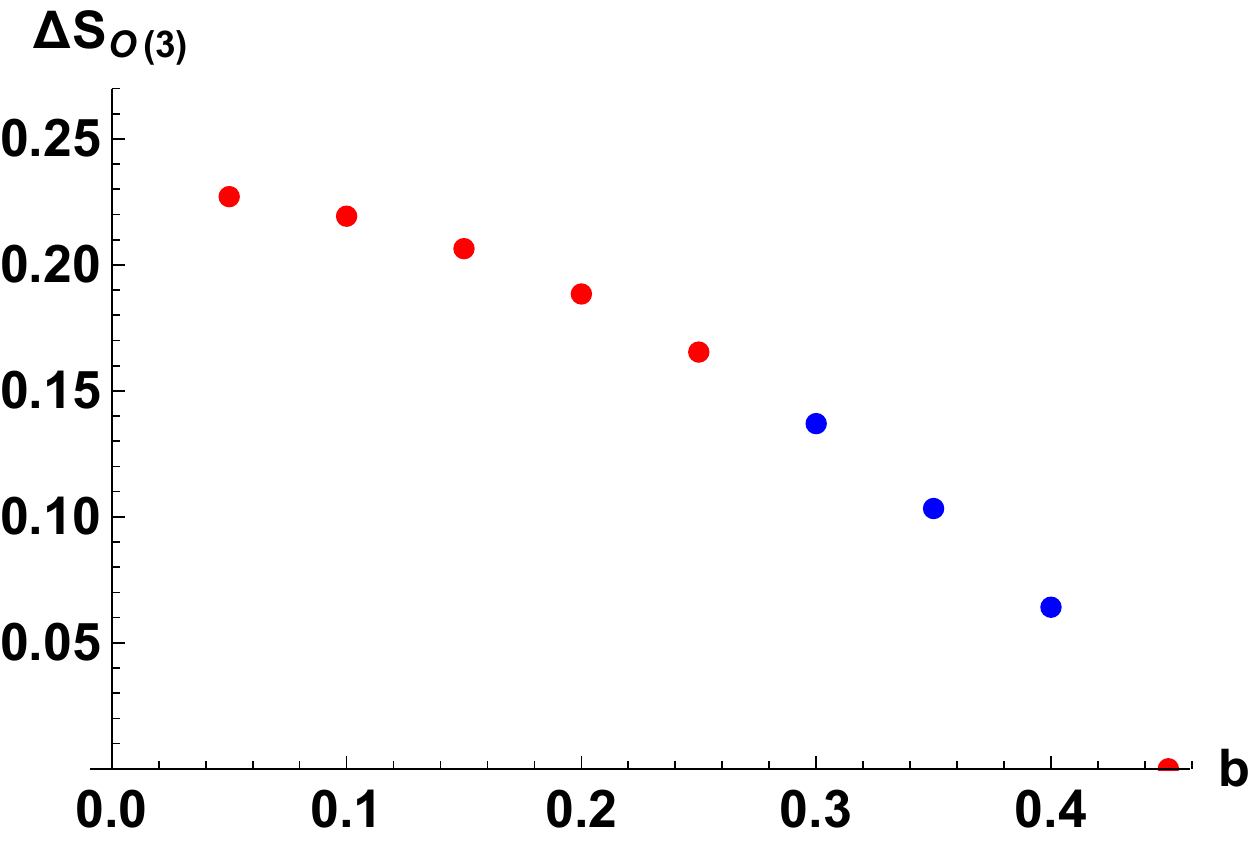} \vspace{0.5cm}
\caption{Numerical plots of $\Delta S_{O(3) } $ changing a value of $b$.
We take $a= 0.04$ and $\tilde{b} = 0.01$ in the left panel and  $\tilde{b} = 0.02$ in the right panel.
The red dots denote the phase transitions with one tunneling
while the blue dots the ones with two sequential tunnelings.}
\label{fig:decay}
\end{center}
\end{figure}

We now estimate the decay rate numerically.
Figure~\ref{fig:decay} shows numerical plots of $\Delta S_{O(3) } $ changing values of $b$.
We take $a= 0.04$ and $\tilde{b} = 0.01, 0.02$ in the left and right panels respectively.
The red dots denote the phase transitions with one quantum tunneling
while the blue dots the ones with two sequential tunnelings.
As the value of $b$ is large and reaches a critical point, the decay rate is significant and the vacuum becomes unstable.

\section{Conclusions and discussions}

Background fluxes are important ingredients in string compactifications. 
If the string landscape is realized in nature, there exist a number of metastable vacua containing such background fluxes.
In this paper, we have pointed out that a baryon-like object can be constructed in a stringy geometry and can be an impurity for the phase transition.
To understand how effectively phase transitions occur in string theory,
it is important to know the catalytic effect induced by this kind of objects. 
The setup we have considered is the large $N$ dual picture of
the D5-brane/anti-D5-brane system discussed in Ref.~\cite{Aganagic:2006ex,Heckman:2007wk}.
Wrapping a D3-brane over a compact 3-cycle,
the presence of background RR fluxes in the cycle induces
the charge of a gauge field on the D3-brane world-volume. Thus, the D3-brane is attached by fundamental strings
and a dibaryon is formed. We also pointed out that a monopole-like D3-brane and the dibaryon make a bound state. 
There are two kinds of transitions in the geometry,
the transition between the $N$ confining vacua and the one reducing the RR flux lines (corresponding to the D5-brane/anti-D5-brane annihilation in the open string picture).
They proceed by wrapping a domain wall D5-brane over a cycle in the internal manifold.
The dibaryon forms a bound state with the domain wall D5-brane
and induces the magnetic flux on the wall.
We find the dibaryon can accelerate the decay of a false vacuum
and threaten the stability itself with some choice of the model parameters.
The proposed new impurity might play a role in understanding the attractor point(s) on the string landscape.

Finally, we comment on possible future directions.
It might be interesting to consider a baryon-like object in the geometry of Ref.~\cite{Kachru:2002gs} and
the catalytic effect on the decay of the false vacuum.
Ref.~\cite{Kachru:2002gs} discussed NS5-brane mediated annihilation between anti-D3 branes and an NSNS 3-form flux
in the Klebanov-Strassler geometry
\cite{Klebanov:2000hb}.
We might be able to construct a new type of bound states with the domain wall NS5-brane. In addition, it would be interesting to study catalytic effects in mestastable vacua in perturbed Seiberg-Witten theories \cite{SW}. The massive monopole can be regarded as an impurity for the semi-classical vacuum decay. These topics are beyond the scope of this paper and we leave them future works.

\section*{Acknowledgment}

The authors are grateful to A.~Kanazawa and F.~Yagi for useful discussions. This work is supported by Grant-in-Aid for Scientific Research from the Ministry of Education, Culture, Sports, Science and Technology, Japan (No. 25800144 and No. 25105011). YN is supported by a JSPS Fellowship for Research Abroad. AK and YO would like to thank Harvard University for their kind hospitality where this work was at the early stage.

%
%

\end{document}